# *Ab initio* modeling of oxygen impurity atom incorporation into uranium mononitride surface and subsurface vacancies

D. Bocharov[1,2,3*], D. Gryaznov[1], Yu.F. Zhukovskii[1], and E.A. Kotomin[1]

[1]*Institute of Solid State Physics, Kengaraga 8, LV- 1063 Riga, Latvia*
[2]*Faculty of Physics and Mathematics, University of Latvia, Zellu 8, LV-1002 Riga, Latvia*
[3]*Faculty of Computing, University of Latvia, Raina blvd. 19, LV-1586 Riga, Latvia*

**Abstract.** The incorporation of oxygen atoms has been simulated into either nitrogen or uranium vacancy at the UN(001) surface, sub-surface or central layers. For calculations on the corresponding slab models both the relativistic pseudopotentials and the method of projector augmented-waves (PAW) as implemented in the *VASP* computer code have been used. The energies of O atom incorporation and solution within the defective UN surface have been calculated and discussed. For different configurations of oxygen ions at vacancies within the UN(001) slab, the calculated density of states and electronic charge re-distribution was analyzed. Considerable energetic preference of O atom incorporation into the N-vacancy as compared to U-vacancy indicates that the observed oxidation of UN is determined mainly by the interaction of oxygen atoms with the surface and sub-surface N vacancies resulting in their capture by the vacancies and formation of O-U bonds with the nearest uranium atoms.

1.  **Introduction**

Uranium mononitride (UN) is considered as a promising fuel for the fast nuclear Generation IV reactors [1]. It has several advantages as compared to the traditional oxide nuclear fuels [2]. The synthesis of UN may be carried out by direct substitution starting from fluorides (*e.g.*, using oxidative ammonolysis of uranium fluorides to uranium nitride [3]). However, oxygen impurities always presented in UN lead to its unwanted pollution and further degradation in air. A number of experiments were performed, in order to understand an influence of oxygen on UN properties and other actinide compounds [4-6]. Various experimental analyses clearly manifested that oxygen contact with the actinide nitride surfaces can result in growth of the oxide [4, 5] and, at initial stages, leads to the formation of oxynitrides [6]. The oxynitrides of varying structure ($UO_xN_y$) can be formed in the reactions of uranium metal with $NO_2$ [7] or by direct current sputtering upon U target in an Ar atmosphere containing admixture of $O_2$ and $N_2$ [8]. This facilitates importance of $UO_xN_y$ for actinide surface studies.

Beginning with 80's [9, 10] a number of *ab initio* calculations on UN bulk [11-16] were performed using formalism of the Density Functional Theory (DFT). It is also worth mentioning that the first-principles calculations on actinide nitride compounds continue attracting great attention, due to improved methods and increasing interest for the fast breeder reactors and for the issues of transmutation of plutonium and minor actinides. Basic bulk properties of actinide nitrides were considered in a few recent studies [17-19], with emphasis on elastic and magnetic properties. The first electronic structure simulations on the perfect UN(001) *surface* and its reactivity towards the molecular and atomic oxygen were performed only recently [20-22]. These studies clearly show that the $O_2$ molecule after adsorption on the UN (001) surface dissociates spontaneously, whereas the newly-produced O atoms demonstrate a strong chemisorption at the surface. For simplicity, we call hereafter O adatoms despite the fact that these are negatively charged ions $O^-$. Additionally, a considerable attention was paid to the static and dynamic properties of primary defects (vacancies) in UN *bulk* [15] which affect the fuel performance during operation and its reprocessing.

* Corresponding author: Fax.:+371 67132778
  E-mail address: bocharov@latnet.lv

Apart the behavior of empty vacancies, the O atom incorporation into vacancies in bulk UN was considered. Its incorporation into the N-vacancies was found to be energetically more favorable as compared to the interstitial sites [23]. However, the solution energy has shown an opposite effect. The migration energy for the O atom via the interstitial sites along the [001] direction is 2.84 eV [23, 24]. Defective UN surface containing both nitrogen and uranium vacancies disposed at different positions within the UN (001) slab has been also discussed in a separate paper [25]. In order to shed more light on the mechanism of UN unwanted oxidation, the incorporation of oxygen impurities into the N- and U- vacancies on the UN(001) surface is focused in this paper.

## 2. Computational details

For the simulation of a defective UN (001) substrate with empty and oxygen-occupied vacancies, the DFT plane wave computer code *VASP 4.6* [26-29] was employed. The *VASP* package is suited for performing first principles calculations based on the DFT approximation when varying the free energy and evaluating the instantaneous electronic ground state at each quantum-mechanical molecular dynamics time step [26, 30]. Ultra-soft pseudopotentials combined with the PAW method [31, 32] were used. Computational procedures implemented in this code [26, 27] foresee the iterative solution of the Kohn–Sham equations based on both residuum-minimization and optimized charge-density mixing routines [28, 29]. The non-local exchange-correlation functional within the Perdew-Wang-91 Generalized Gradient Approximation (PW91 GGA) [33] and the relativistic pseudopotentials for 78 U core electrons (with $6s^2 6p^6 6d^2 5f^2 7s^2$ valence shell), as well as 2 both N and O core electrons (with $2s^2 2p^3$ and $2s^2 2p^4$ valence shells, respectively) were applied in the current study. The Monkhorst-Pack scheme [34] with a 8×8×1 *k*-point meshes in the Brillouin zone (BZ) was used while the cut-off energy was set to 520 eV. It became common in last years to use the so-called GGA+*U* approach to such strongly correlated systems as actinides (*e.g.*, $UO_2$ [35] and references therein). On the other hand, our test calculations with reasonable *U*-parameters have shown that the relative variation in the defect formation energies may be ~10 per cent which does not affect main trends and conclusions of the present study. This is why the standard GGA approximation was used which is important for a comparison with previous calculations of defect energies in the UN bulk [15].

For the symmetric UN(001) substrate possessing the *fcc* rock-salt structure, a slab model was employed. It consists of 5, 7 or 9 atomic layers and containing regularly alternating U and N atoms. The 2D slabs are separated by a vacuum gap of sixteen interlayer distances (38.88 Å) in the *z*-direction. This inter-slab distance is large enough to exclude the spurious interaction between the slabs. Our calculations supposed the supercells with 2×2 and 3×3 extensions of translation vector for the (001) surface of UN. Both empty as well as oxygen-occupied N- and U-vacancies were disposed in the surface, subsurface and central layers of 2D slab. Due to the presence of mirror layers in the symmetric slabs, one can consider the two-sided symmetric arrangement of defects (Fig. 1), except for the central mirror plane, thus, minimizing the computational expenses. Figure 1 also shows the oxygen-occupied N vacancies with a 2×2 and 3×3 periodicity disposed on the surface layer. The lattice constant of UN slabs is fixed at 4.87 Å, taken from the lattice relaxation of the UN bulk [20]. In all the calculations, the structural optimization within the supercell of fixed linear dimensions was performed, using the standard procedure of total energy minimization. Our test calculations have shown that the ferromagnetic (FM) phase is energetically slightly more favorable for UN *slabs* than the anti-ferromagnetic (AFM) one. The spin magnetic moment was allowed to relax in all the calculations for the FM spin arrangements on the uranium sub-lattice.

### 3. Incorporation and solution energies

The energy balance for the incorporation of an O atom into a vacancy can be characterized by *the incorporation energy $E_I$* suggested by Grimes and Catlow [36] in the shell model calculations on fission products in $UO_2$:

$$E_I = E_{O\_inc}^{N(U)} - E_{vac}^{N(U)} - E_O \tag{1a}$$

for the O atom incorporated into the N- and U-vacancy disposed in the central atomic layer and

$$E_I = \frac{1}{2}(E_{O\_inc}^{N(U)} - E_{vac}^{N(U)} - 2E_O) \tag{1b}$$

for the same incorporation in the surface or sub-surface layers. Here $E_{O\_inc}^{N(U)}$ is the total energy of the supercell containing the O atom at either the N- or U-vacancy ($E_{O\_inc}^{N(U)} < 0$), $E_{vac}^{N(U)}$ the energy of the supercell containing an unoccupied (empty) vacancy, and $E_O$ half the total energy of isolated $O_2$ molecule in the triplet state. It is defined by the oxygen chemical potential at 0 K. Since the value of $E_I$ describes the energy balance for the incorporation into pre-existing vacancies, it has to be negative for energetically favorable incorporation processes.

To take into account the total energy balance, including the vacancy formation energy $E_{form}$ in the defect-free slab, the solution energy [36] was defined as:

$$E_S = E_I + E_{form}. \tag{2}$$

where $E_{form}$ is the formation energy of N- or U- vacancy in the slab, calculated as

$$E_{form} = E_{vac}^{N(U)} + E_{atom}^{N(U)} - E^{UN} \tag{3a}$$

for a defect in the central atomic layer of the slab and

$$E_{form} = \frac{1}{2}(E_{vac}^{N(U)} + 2E_{atom}^{N(U)} - E^{UN}) \tag{3b}$$

for a defect in the surface or sub-surface layer. Here $E^{UN}$ is an energy of the defectless relaxed slab, and $E_{atom}^{N(U)}$ can be defined as chemical potentials of N or U atom, which is, in general, a function of temperature and nitrogen partial pressure. The chemical potential of nitrogen at 0 K is defined by the total energy of $N_2$ molecule, *i.e.*, $m_{N_2} = \frac{1}{2} E_{tot}[N_2]$, while the chemical potential of U atom at 0 K can be estimated as the total energy, *per* atom, for metallic uranium in its low-temperature ortho-rhombic α-phase: $m_{\alpha\text{-}U} = \frac{1}{2} E_{tot}[\alpha\text{-}U]$. The co-factor of ½ in Eqs (1b, 3b) as well as multiplication of $E_O$ and $E_{atom}^{N(U)}$ by 2 in the same equations appears due to the symmetric arrangement of incorporated O atoms.

More details on calculations of unoccupied N- and U- vacancies and parameters of calculated of N$_2$ and α-U are given in Ref. [25]. It is worth mentioning, however, that use of the standard O pseudopotential in our VASP calculations gave good bond length of 1.23 Å for the O$_2$ molecule but considerable overestimate of the binding energy (6.79 eV *vs.* the experimental value of 5.12 eV). Several corrections were suggested in the literature how to take into account this serious DFT shortcoming [37, 38]. Thus, the calculated defect formation and solution energies may be corrected by ~1 eV (its impact is also discussed below).

## 4. Results and discussion

The calculated O atom incorporation into the N-vacancies at the UN(001) surface is energetically favorable since both values of $E_I$ and $E_S$ are strictly negative (Table 1), thus, being in favor of both creating the N-vacancy and adsorbing the O atom from air. Also, $E_I$ decreases by ~0.4 eV (becomes more negative) within the slab as compared to the surface layer, whereas $E_S$ is smallest for the N-vacancy just on the surface layer. Contrary, in case of the U-vacancies, the values of $E_I$ calculated for the surface and central layers are found to be close to zero. The sub-surface layer is characterized by $E_I$ which is ~1 eV smaller than that for the surface and central layers. Our results indicate importance of oxynitride formation. However, $E_S$ is positive and increases for O atoms in the the U-vacancy and the slab centre. Note that the energies in Table 1 do not include the corrections discussed above for the O atoms. However, such corrections may lead to $E_I$ (or $E_S$) increased by ~1 eV and, as a result, more positive $E_I$ for the U-vacancy. Table 1 also indicates that solution of the oxygen atoms is energetically more favorable at the surface layers than in the slab. As the supercell size increases (the 3x3 extension in Table 1), both $E_I$ and $E_S$ values decrease whereas the slab thickness has no such clear effect. Nevertheless, the U-vacancy appeared to be most sensitive to the supercell size which is related to spurious interactions between the periodically repeated defects. The $E_I$ as well as $E_S$ values may be reduced by 0.15 eV at the average in this case.

Table 1 allows us to analyze also the averaged spin density of U atoms $m_{av}^U$ for different morphologies of defective UN (001) surfaces with incorporated O atoms. Analogously to defective UN surface with empty vacancies [25], $m_{av}^U$ decreases with a number of layers in the slab for both types of the vacancies (except for the O atom incorporated into the U-vacancy in the surface layer). It is also seen that $m_{av}^U$ is higher in the surface layer for the N-vacancy than for the U-vacancy. The sub-surface and central layers are characterized by similar $m_{av}^U$ for both the vacancies. Interestingly, the effective charge $q_{eff}$ of O atoms is also higher for the N-vacancy and inside the slab. In the case of U-vacancy, however, $q_{eff}$ decreases by almost 0.3 e. The same effect is also seen for the N atoms their effective charge is smaller when the O atom occupies the U-vacancy (not shown here). The overall picture suggests prevalence of the covalent bonding between different species in the system.

Large concentrations of defects (25% for the 2×2 extension in Table 1) causes certain finite-size effects which can be illustrated using the 2D difference electron density redistributions $\Delta\rho(\mathbf{r})$. These redistributions are shown for the O atoms incorporated into the N-vacancies at the surface (Fig. 2) and central layers (Fig. 3). Inside the 5-layer slab, a presence of the two symmetrically positioned defects induces their interaction (visible in charge redistribution across a slab in Fig. 2a). An increase of the slab thickness reduces this effect (Fig. 2c). If the supercell size is decreased (the 2×2 extension, Fig. 2b) an additional electron density parallel to the surface layer is observed between the defects. Similar effects are also observed for redistributions of the electron density around defects in the mirror planes (Fig. 3). A perfect spherical negative charge redistribtion is observed around the O atom in the U-vacancy in the central plane (not shown here). The effect of supercell size in this case is similar to that discussed for the N-vacancy. However, in the case of surface vacancy more electron density is seen between the O atom and neighbouring N atoms in the sub-surface layer, in a comparison to the N-vacancy. Thus, the effect of slab thickness also may not be underestimated in this case.

In Figure 4, the total and projected density of states (DOS) is shown for the 7-layer defective UN(001) surface with the O atom incorporated into the N-vacancy. The system remains conducting throughout all the calculations with the significant contribution from the U *5f* states at the Fermi level similar to pure UN bulk [20]. The appearance of the specific O *2p* band with the energy peak at –6 eV is observed. When comparing the DOS for the O atoms incorporated into the N–vacancies, a noticeable shift of the O *2p* band (by about -1.0 eV) allows one to distinguish the surface layer from the internal layers. Moreover, in the case of the surface layer, this band considerably overlaps with the N *2p* band, partly mixed with the U *5f* states (similar effects happen with the $O_2$ molecule atop the surface U atom [22]). Contrary, the O *2p* band remains quasi-isolated from the other bands (analogously to the O atom incorporated into the N-vacancy in UN bulk [23]). Note that position of the N *2p* band is insensitive to the presence of O atoms and lies within energy range of -6 and -1 eV.

**4. Conclusions**

Summing up the results obtained in this and our recent studies, the following stages for reactivity of oxygen positioned atop the UN(001) surface could be suggested: (*i*) spontaneous breaking of the $O_2$ chemical bond after molecular adsorption [22], (*ii*) location of the two newly formed O adatoms atop the surface U atoms [21], (*iii*) incorporation of O adatoms in pre-existing surface N vacancies (as a result of vacancy surface diffusion), (*iv*) incorporation of O atoms in existing subsurface N vacancies as a result of inter-lattice diffusion. This explains an easy UN oxidation observed in air.

The formation of oxynitrides [8] near the UN(001) surface is proposed, which can be caused by diffusion of the oxygen atoms within the interlayers of uranium nitride with further capture by nitrogen vacancies, thus, resulting in their stabilization due to formation of the chemical bonds with the nearest uranium atoms. The relevant effects of the electronic charge redistribution were analyzed. They demonstrate a quite local nature of the density perturbation caused by the incorporated O atoms. The analysis of density of states shows both overlapping of the O 2$p$ states with the N 2$p$ states at initial stages of oxidation (*surface incorporation*) and separation of the O 2$p$ states from other bands in the case of deeper positioned oxygen atoms (*sub-surface penetration*). The results of this analysis could be used for the interpretation of the experimental ultraviolet photoelectron spectra for uranium oxynitrides [8].

**Acknowledgements**

This study was partly supported by the European Commission FP7 project *F*-BRIDGE and ESF project No. 2009/0216/1DP/1.1.1.2.0/09/APIA/VIAA/044. The corresponding author gratefully acknowledges also the doctoral studies support by the European Social Fund. The authors kindly thank R.A. Evarestov, P. Van Uffelen and V. Kashcheyevs for a numerous fruitful discussions. The technical assistance of A. Gopejenko and A. Kuzmin was the most valuable.

**Figures captions**

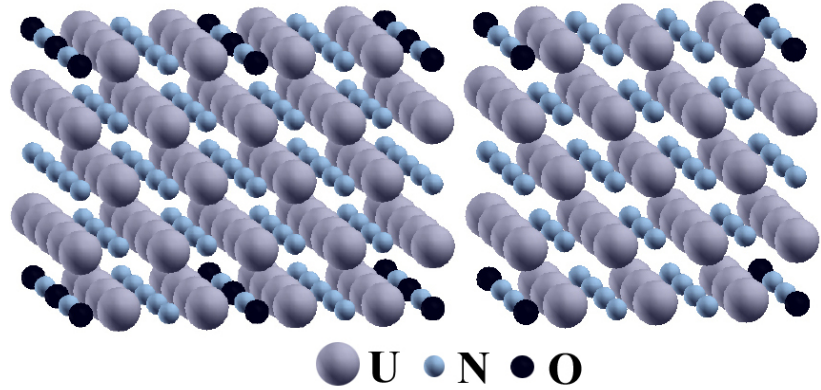

*Figure 1.* The symmetrical 5-layer UN(001) slab with a 2×2 (left) and 3×3 (right) periodicity of the oxygen atoms incorporated into the surface N-vacancies.

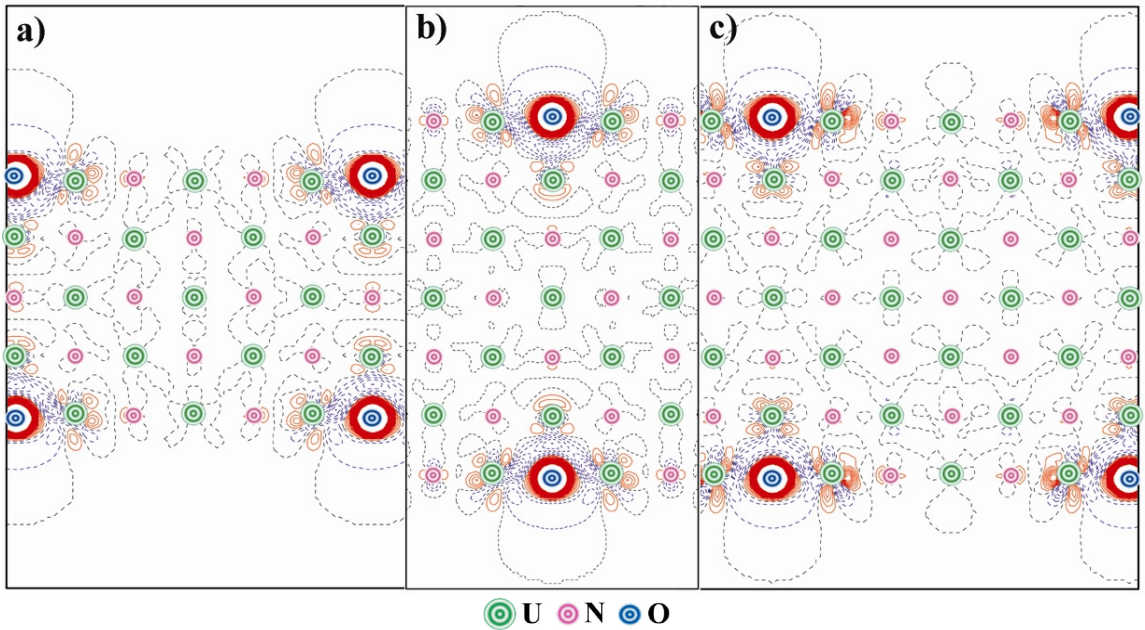

*Figure 2* (Color online). The 2D sections of the electron charge density re-distributions $\Delta\rho(\mathbf{r})$ around the O atoms incorporated into the surface N-vacancies of the five- and seven-layer UN(001) slabs with 2×2 and 3×3 supercell extensions. $\Delta\rho(\mathbf{r})$ are defined as the total electron density of the O-containing defected surface minus a superposition of the electron densities of the surface containing the N-vacancies and the O atom in the regular positions on the surface. a) 3×3 periodicity of the oxygen atoms upon the five-layer slab, b) 2×2 periodicity of the oxygen atoms upon the seven-layer slab, c) 3×3 periodicity of the oxygen atoms upon the seven-layer slab. Solid (red) and dashed (blue) isolines correspond to positive and negative electron density, respectively. Dashed black isolines correspond to the zero-level (For interpretation of the references to colour in this figure legend, the reader is referred to the web version of this article). Isodensity increment is 0.25 e a.u.$^{-3}$.

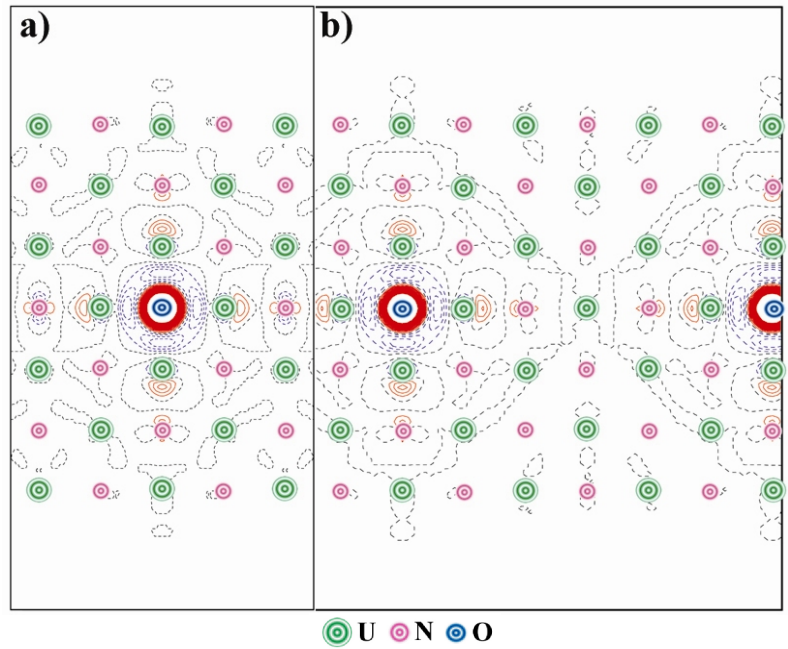

*Figure 3* (Color online). The 2D sections of *Δρ*(**r**) around the O atoms incorporated into the N-vacancies disposed in central layer of the seven-layer UN(001) slabs with (a) 2×2 and (b) 3×3 supercell extensions. Other details are given in caption of Figure 2.

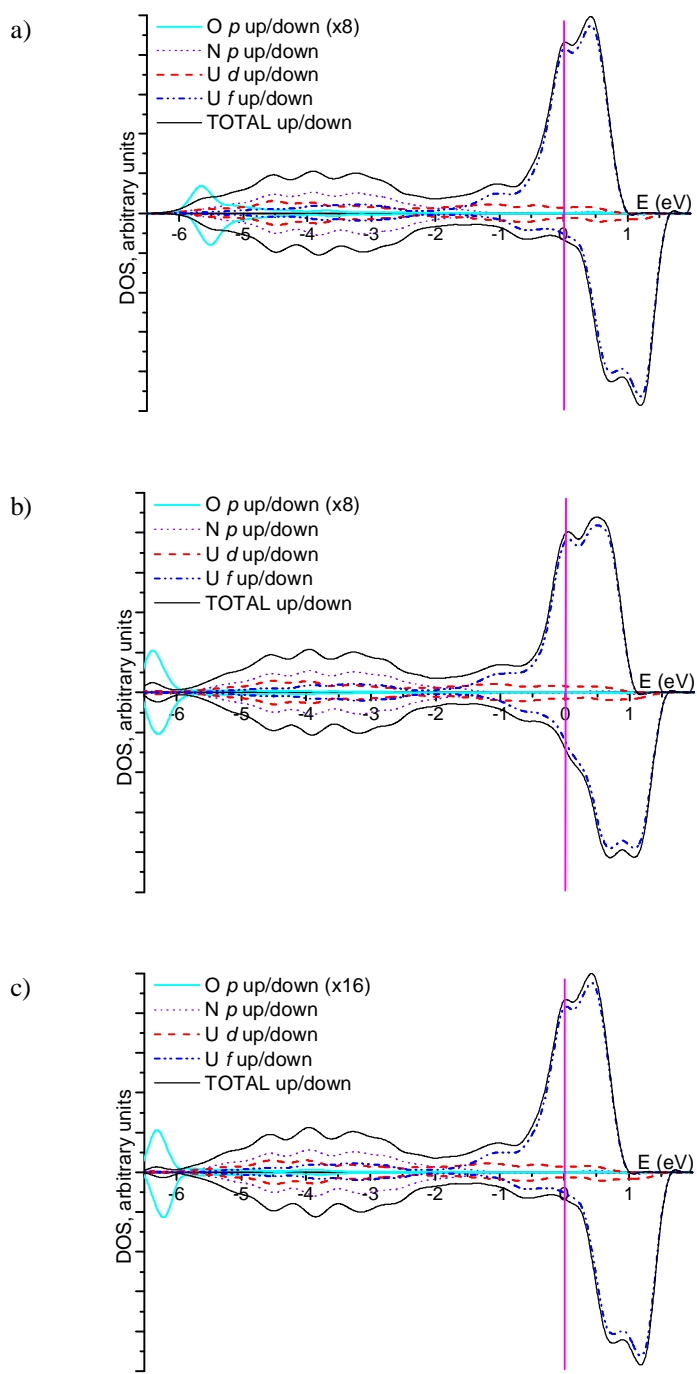

*Figure 4* (Color online). The total and projected DOS for three positions of O atoms incorporated into the N vacancies with a 3×3 periodicity across the 7−layer UN(001) slab: a) surface layer, b) sub-surface layer, c) central layer. The O 2p peaks have been normalized to the same value, *i.e.*, these were multiplied by a factor of 8 and 16 for vacancies in the surface (subsurface) and central layers, respectively (please see figure labels), A convolution of individual energy levels has been plotted using the Gaussian functions with a half-width of 0.2 eV.

Table 1. Incorporation ($E_I$), solution ($E_S$) energies in eV, average spin magnetic moments of U atoms $m_{av}^U$ in $\mu_B$ and effective charge of O atoms in e⁻ for O incorporation into the UN(001) surface. The reference states for calculating the incorporation and solution energies are the chemical potentials of O, N and U calculated for $O_2$, $N_2$ molecules and α-U, respectively.

| Layer | Supercell size | Number of atomic layers in slab | N vacancy | | | | U vacancy | | | |
|---|---|---|---|---|---|---|---|---|---|---|
| | | | $E_I$ (eV) | $E_S$ (eV) | $m_{av}^U$ ($\mu_B$) | $q_{eff}$(e⁻) | $E_I$ (eV) | $E_S$ (eV) | $m_{av}^U$ ($\mu_B$) | $q_{eff}$(e⁻) |
| Surface | 2x2 | 5 | -6.173 | -2.473 | 1.65 | -1.36 | -0.339 | 1.120 | 1.16 | -0.98 |
| | | 7 | -6.181 | -2.476 | 1.49 | -1.36 | -0.855 | 0.583 | 1.36 | -1.03 |
| | | 9 | -6.188 | -2.479 | 1.41 | -1.36 | -0.943 | 0.493 | 1.31 | -1.06 |
| | 3x3 | 5 | -6.122 | -2.481 | 1.60 | -1.37 | -0.683 | 0.654 | 1.48 | -1.05 |
| | | 7 | -6.126 | -2.480 | 1.46 | -1.36 | -1.073 | 0.230 | 1.38 | -1.08 |
| Subsurface | 2x2 | 5 | -6.314 | -2.068 | 1.64 | -1.42 | -1.856 | 1.284 | 1.66 | -1.10 |
| | | 7 | -6.419 | -2.090 | 1.49 | -1.40 | -1.823 | 1.297 | 1.45 | -1.10 |
| | | 9 | -6.417 | -2.091 | 1.41 | -1.40 | -1.823 | 1.271 | 1.38 | -1.10 |
| | 3x3 | 7 | -6.428 | -2.093 | 1.46 | -1.39 | -2.012 | 1.000 | 1.43 | -1.10 |
| Central (mirror) | 2x2 | 7 | -6.611 | -2.180 | 1.47 | -1.42 | 0.736 | 3.923 | 1.44 | -0.89 |
| | | 9 | -6.608 | -2.192 | 1.39 | -1.38 | 0.669 | 3.838 | 1.38 | -0.90 |
| | 3x3 | 7 | -6.599 | -2.182 | 1.45 | -1.42 | 0.317 | 3.378 | 1.47 | -0.94 |